\documentclass[12pt]{article}
\usepackage{epsfig}
\begin{document}

\newcommand{\refp}[1]{(\ref{#1})}
\newcommand{\beq}{\begin{equation}}
\newcommand{\eeq}[1]{\label{#1} \end{equation}}
\newcommand{\beqar}{\begin{eqnarray}}
\newcommand{\eeqar}[1]{\label{#1} \end{eqnarray}}
\newcommand{\ket}[1]{| {#1} \rangle }
\newcommand{\bra}[1]{\langle {#1} | }

\begin{center}

\begin{Large}

On the Motion of Zeros of Zeta Functions\\[3ex]

\end{Large}

Hans Frisk and Serge de Gosson$~^{*}$ \\[1ex]

Department of Mathematics, 
University of V\"axj\"o, S-351 95 V\"axj\"o, Sweden\\[2ex]

\end{center}

\begin{quotation}
The motion in the complex plane of the zeros to various 
zeta functions is investigated numerically.
First the Hurwitz zeta function is considered and an accurate
formula for the distribution of its zeros is suggested. Then
functions which
are linear combinations of different Hurwitz zeta functions, and
have a symmetric distribution of their zeros with respect to
the critical line, are examined. Finally the existence of the
hypothetical
non-trivial Riemann zeros with $Re~s\neq1/2$ is discussed.
 \\[75ex]
\end{quotation}
*) e-mail addresses: hans.frisk@msi.vxu.se and serge.degosson@msi.vxu.se
\clearpage
\section{Introduction} 
The Riemann zeta function $\zeta(s)$ \cite{edw,tit,vor,ivi} and its zeros
is a source of inspiration in
the field of quantum chaos \cite{berry}.
It is well
known that the imaginary parts of the zeros to $\zeta$
on the critical line $Re~s=1/2$ 
have some
striking similarities with a spectra of an Hermitian
operator with time reversal symmetry broken \cite{metha}.
The famous Riemann hypothesis states that in fact all non-trivial
zeros of $\zeta$ lie on the critical line and the search
for the Hermitian operator, and its classical counterpart, has been
intensified recently \cite{ext}.\\[1ex] 
With help of powerful computers the Riemann 
zeros 
have been explored in great detail \cite{odlyz}. In this paper we present
much more modeste numerical experiments for the
zeros of Hurwitz zeta function \cite{vor,apost}
\beq
\zeta (s,\alpha )=\sum_{n=0}^{\infty} \frac{1}{(n+\alpha)^s},~~
0< \alpha \leq 1,~\sigma>1,~s=\sigma + it,  
\eeq{hdef}
where $\zeta(s,1)=\zeta(s)$, and for linear combinations of them,
\beq
\Psi(s)~=~\sum_{l=1}^{m}~c_{l}~\zeta (s,\alpha_{l}).
\eeq{lcomb} Important examples of $\Psi(s)$ are the
L-functions \cite{vor,apost,rum} for which $\alpha_{l}=l/m$
and $c_{l}=\chi(l)/m^{s}$, $\chi(l)$ are Dirichlet characters
modulo $m$. 
The motivation for the study is that it is often very
useful to study how a system behaves as function of a 
parameter, see e.g. chapter 6 in \cite{arn1}. 
We will, inspired by the similarities mentioned above, 
use the word "spectra" for the imaginary parts 
of the zeros, not only for $\zeta(s,1)$
but also for $\zeta(s,\alpha)$ and $\Psi(s)$.  
In sect. 2 we consider the spectra for 
$\zeta (s,\alpha)$ as function of $\alpha$
while we in sect. 3 discuss the motion of the zeros for some
$\Psi(s)$ with $m=5$, and symmetric distribution of the zeros.
In the latter case the coefficients, $c_{l}$, depend on one or two
parameters and it is the motion with respect to their variations
that is studied. 
Bombieri and Hejhal has shown that in 
the limit of large $t$ almost
all zeros for these symmetric functions are simple and lie on the
critical line \cite{hejhal}. Finally in sect. 4 some results on
symmetric zeta functions with $m>5$ are presented together with some
speculations on the existence of non-trivial Riemann zeros with $Re~s\neq1/2$. 
The numerical calculations presented
are done with Mathematica on a Sun Ultra
machine.
\\[2ex]
\section{Hurwitz zeta function}
In this section we focus  on the motion of the zeros of $\zeta(s,\alpha)$
with respect to $\alpha$. Since along a zero $d\zeta = \zeta'~ds +
\partial\zeta/\partial\alpha~d\alpha=0$ we get the following equation
for the motion of a zero at $s=z$
\beq
\frac{dz}{d\alpha}=-\frac{\partial\zeta(z,\alpha)}{\partial\alpha}\cdot
\frac{1}{\zeta'(z,\alpha)}=\frac{z\cdot\zeta(z+1,\alpha)}{\zeta'(z,\alpha)}.
\eeq{zmot}
The second equality in (3) is obtained by derivating (1) with
respect to $\alpha$. For the extension of $\zeta(s,\alpha)$ to the
whole complex plane by analytical continuation see \cite{vor,apost}.
Note that when  $\alpha$ varies zeros can merge to multiple
zeros or go away to infinity but never disappear or be created.
At multiple zeros $\zeta'=0$ so the equation of motion must be
modified, see sect. 3. 
A spectra for the Hurwitz zeta function can now be 
obtained by regarding $t=Im~z(\alpha)$ as
an eigenvalue. To get a spectra at $\alpha=1$ with unit mean
spacing one
has to use scaled eigenvalues $N(t)$. With $N(T)$ we denote the
number of zeros in the range $0< Im~z \leq T$,
$0\leq Rez \leq 1$, 
and the following asymptotic formula \cite{edw,tit,vor}, 
\beq
N(T)=\frac{T}{2\pi}Log(\frac{T}{2\pi})-\frac{T}{2\pi},
\eeq{nt1}
is used.
To obtain a spectra with approximately unit mean spacing 
in the whole region
$0<\alpha\leq1$ we suggest that (4) should be modified in the
following way 
\beq
N(T,\alpha)=\frac{T}{2\pi}Log(\frac{T}{2\pi})-\frac{T}{2\pi}
-\frac{T}{2\pi}Log(\alpha),
\eeq{nt2}
where the zeros now can be located outside the critical strip.
Numerical experiments in the range $10^{-10}\leq\alpha\leq$1
, $0<T\leq~10^{4}$ indicate that (5) is very accurate but 
we have not been able
to  prove it. It is certainly correct for
$\alpha=1/2$ since
\beq
\zeta(s,1/2)=(2^{s}-1)\cdot\zeta(s,1).
\eeq{alf12}
The zeros of $2^{s}-1$ have $\sigma=0$ with spacings
$\Delta t=\frac{2\pi}{Log(2)}$ and their number 
up to $Im~z=T$ is given by the last term in (5). The formula also seems to be accurate
for  L-functions if $m=p$, $p$ is a prime number, and
$\alpha=1/p$ is used in (5), see also \cite{ivi}.

In fig. 1 a part of the spectra for the Hurwitz zeta
function is presented. The 30:th to the 43:th zero
in the range $10^{-2}\leq\alpha\leq1$ are shown. 
We can observe that the spectra share some 
properties with a generic quantal spectra  since there are many avoided
crossings but also degeneracies. Note that these degeneracies
are only in $t$, not in $\sigma$. 
However, we see also that the crossings occur in a somewhat peculiar way. They
take place when $N(t(\alpha))$ for one of the zeros is strongly  
decreasing with increasing $\alpha$ and then usually more crossings
occur for nearlying  $\alpha$-values. This effect is much more pronounced
higher up in the spectra.
It can also be seen from fig. 1 that the spectra becomes
more equidistant when $\alpha\rightarrow0$ . 
This can be understood qualitatively since
$\zeta(s,\alpha)\approx\frac{1}{\alpha^{s}}+\zeta(s,1)$ for
small $\alpha$-values. 
The zeros seem thus to move towards $s=0$ when $\alpha\rightarrow 0$.
From the analytical continuation of (1) to the whole complex plane
we know that $\zeta(s,1)$ has trivial zeros at $s=-2,-4,-6,....$
From (3) we expect that these zeros move to the right on the $\sigma$-
axis when $\alpha$ decreases and eq. (6) gives that these zeros
have moved two units at $\alpha=0.5$. When $\alpha\rightarrow0$ the
zeros approach $s=1,0,-2,-4,-6,...$.

\section{Symmetric functions $\Psi(s)$ for $m=5$}
Symmetric functions here means functions of the
type given by eq. 2 for which the non-trivial zeros are located
symmetrically around $\sigma=1/2$, i.e., if $\Psi(z)=0$
also $\Psi(1-\bar{z})=0$. The symmetric $\Psi(s)$ which also are
L-functions can be expressed as
Euler products and then it is straightforward
to show that $\Psi(s) \neq0$ when $\sigma>1$ \cite{vor}. The three functions
of this type with the lowest $m$-values are
$\zeta(s,1)$,
$\frac{1}{3^s}(\zeta(s,1/3)-\zeta(s,2/3))$ and
$\frac{1}{4^s}(\zeta(s,1/4)-\zeta(s,3/4))$.
\\[1ex]
For $m\geq5$ opens up the possibility of constructing
symmetric $\Psi(s)$ which are linear 
combinations of L-functions. For these linear combinations
there is generally no Euler product and zeros can appear
outside the critical strip.
For $m=5$ we can for example make the following ansatz
\beq
\Psi_{5o}(s,\beta)=\frac{1}{5^s}(\zeta(s,1/5)-\zeta(s,4/5)+
\beta~(\zeta(s,2/5)-\zeta(s,3/5))),
\eeq{5o}
,$o$ denotes odd since $c_{l}=-c_{m-l}$ in (7).
Following the lines at p. 212 in \cite{vor} a
condition
for $\Psi_{5o}(s,\beta)$ being symmetric can be obtained,
\beq
sin\frac{4\pi}{5}-\beta~sin\frac{2\pi}{5}=\bar{\beta}~(
sin\frac{2\pi}{5}+\beta~sin\frac{4\pi}{5}).
\eeq{c5}
The two odd
L-functions with $m=5$ have $\beta=\pm i$. In fig. 2
a part of 
the spectra for $\Psi_{5o}(s,\beta)$, with $\beta$ lying on the circle 
(8), is shown. To obtain this figure the equation of motion (3) is
used but with $\alpha$ and $\zeta(s,\alpha)$ replaced by $\beta$
and $\Psi_{5o}(s,\beta)$. 
In contrast to the spectra for
the Hurwitz zeta function double zeros now
appear frequently and in the figure four of them
can be seen. It happens when two zeros meet on the critical
line and  move outside $\sigma=1/2$ and then again when
these two zeros join on the critical line. At multiple
zeros $\Psi_{5o}'(z,\beta)=0$ so $\frac{dz}{d\beta}$ becomes infinite there.
To overcome this difficulty a small complex constant can be added
to the parameter. For each turn on the circle (8) in the counter-
clockwise sense a zero from the lower half-plane moves to the
upper half-plane and this is why most zeros in fig. 2 move upwards.
It is supposed, 
but not proven, that for the
two L-functions at $\beta=\pm i$ all non-trivial zeros lie on the
critical line. It seems hard to gain insight to that problem from
the study of this one-parameter family of symmetric functions. It
becomes more interesting when we now consider to a two-parameter family.

In the same way as above we can try to construct
even, i.e. $c_{l}=c_{m-l}$, and symmetric functions with $m=5$,
\beq
\Psi_{5e}(s,\beta)=\frac{1}{5^s}(\zeta(s,1/5)+\zeta(s,4/5)+
\beta~(\zeta(s,2/5)+\zeta(s,3/5))).
\eeq{5e}
However, it now turns out that besides the condition, see again p.212 in \cite{vor}
\beq
cos\frac{4\pi}{5}+\beta~cos\frac{2\pi}{5}=\bar{\beta}~(
cos\frac{2\pi}{5}+\beta~cos\frac{4\pi}{5})
\eeq{c5e}
$\beta$ must also fulfill 
$1+\beta=0$ so $\beta=-1$ is the only possibility,
which is a L-function. To make a continous change of
the spectra, as for $\Psi_{5o}$, we can take a linear
combination of $\Psi_{5e}$ with $\zeta(s,1)$, 
\beq
\Psi_{e}(s,\beta,\gamma)=\Psi_{5e}(s,\beta)+\frac{\gamma}{5^s}\zeta(s,1).
\eeq{5e2}
The symmetry requirement leads to the following 
equations for the coefficients $\beta$ and $\gamma$
\beqar
cos\frac{4\pi}{5}+\beta~cos\frac{2\pi}{5}+\frac{\gamma}{2}=
\bar{\beta}~(cos\frac{2\pi}{5}+\beta~cos\frac{4\pi}{5}+\frac{\gamma}{2} )&\nonumber\\
1+\beta+\frac{\gamma}{2}=\bar{\gamma}~(cos\frac{2\pi}{5}+\beta~cos\frac{4\pi}{5}+\frac{\gamma}{2})
.&
\eeqar{c5e2}
For $\beta=1$ the first equation is fulfilled for any $\gamma$ while the
second equation gives  $\gamma=1 + \sqrt{5}~e^{i\theta}$ where
$\theta$ is an angle between $0$ and $2\pi$. Besides
the symmetric functions given by (12) there is one more possibility,
namely $\beta=\gamma=1$. At this point, where 
$cos\frac{2\pi}{5}+\beta~cos\frac{4\pi}{5}+\frac{\gamma}{2}=0$,
$\Psi_{e}(s,1,1)$ reduces to $\zeta(s,1)$ but it is not possible to
continously vary this spectra since 
$\beta=\gamma=1$ is not a solution to (12). On the circle 
$\beta=1,\gamma=1 + \sqrt{5}~e^{i\theta}$ we have
\beq
\Psi_{e}(s,\theta)=(1 +\frac{e^{i\theta}}{5^{s-1/2}})\zeta(s,1).
\eeq{onci}
When $\gamma$ moves counter-clockwise along the circle the 
zeros of $1 +\frac{e^{i\theta}}{5^{s-1/2}}$ (in the following denoted
as trivial),
separated by $\Delta t=2\pi/Log(5)$, 
move upwards with $\frac{dt}{d\theta}=1/Log(5)$ while the 
zeros of $\zeta(s,1)$ remain fixed.\\[1ex]
Let us now put 
$\beta=1+\epsilon~e^{i\varphi}$ and consider when $\beta$ moves on a circle
with radius $\epsilon$ around $\beta=1$ in the counter-clockwise sense.
Then eq. 12 gives 
\beq
\gamma=1 + \sqrt{5}~e^{i2\varphi} + \epsilon\frac{1+\sqrt 5}{2}
e^{i\varphi},
\eeq{modci}
$0 \leq \varphi < 2 \pi$.
With these coefficients we get
\beq
\Psi_{e}(s,\epsilon,\varphi)=(1 +\frac{e^{i2\varphi}}{5^{s-1/2}})
\zeta (s,1) + \epsilon\frac{e^{i\varphi}}{5^{s}}~g_{5}(s),
\eeq{onmod}
where
\beq
g_{5}(s)=
\frac{1+\sqrt 5}{2}
\zeta (s,1) + 
\zeta(s,2/5)+\zeta(s,3/5).
\eeq{g5s}
Compared to (13), 
$\Psi_{e}(s,\epsilon,\varphi)$ in (15) have
no trivial zeros in the critical strip. For small values of $\epsilon$ the two
cases must, however, be similar, see fig. 3. The
difference is that when an upsloping "almost trivial" 
zero meet an "almost Riemann"
zero the two zeros bifurcate out in the
complex plane or they interact and exchange character on
the critical line.\\[1ex] 

The equation of motion for a simple zero,
$z_{\epsilon}(\varphi)$, to (15) for a fixed positive $\epsilon$ is
\beq
\frac{dz_{\epsilon}}{d\varphi}=-\frac{\partial\Psi_{e}(z_{\epsilon},\epsilon,\varphi)}
{\partial\varphi}\cdot
\frac{1}{\Psi_{e}'(z_{\epsilon},\epsilon,\varphi)}=-i~\zeta(z_{\epsilon},1)~
(\frac{e^{i2\varphi}}{5^{z_{\epsilon}-1/2}}-1)\cdot
~\frac{1}{\Psi_{e}'(z_{\epsilon},\epsilon,\varphi)}.
\eeq{zmot2}

The fix points of this dynamical system are
the Riemann zeros. On the critical line, where $Re~z_{\epsilon}=1/2$,
turning points also appear.
For large values of $\epsilon$ the second term in (15) dominates
and for the symmetric function  
$g_{5}(s)$ zeros outside the critical line
appear frequently. If 
$z_{\infty}$ denotes a simple zero of $g_{5}(s)$ and if
$\Delta z(\varphi)
= z_{\epsilon}(\varphi)-z_{\infty}$ the equation of
motion (17) becomes to lowest order
\beq
\Delta z'(\varphi)=
i~\Delta z~
(\frac{e^{i2\varphi}}{5^{z_{\infty}-1/2}}-1)/
(\frac{e^{i2\varphi}}{5^{z_{\infty}-1/2}}+1)
\eeq{zmot3}
This linearised system gives rise to straight line motion if
$Re~z_{\infty}=1/2$ and cycles when $Re~z_{\infty}\neq1/2$. These cycles
are counter-clockwise for $Re~z_{\infty}<1/2$ and clockwise for
$Re~z_{\infty}>1/2$. Thus, for large values of 
$\epsilon$ the zeros of $\Psi_{e}(s,\epsilon,\varphi)$
near the zeros of $g_{5}(s)$ perform vibrational or circulating motion.
However, far outside the critical strip
there is still an upward flow of almost trivial zeros
since $\Psi_{e}(s,\epsilon,\varphi)
~\approx~1 + \epsilon\frac{e^{i\varphi}}{2^{s}}$ for large
values of $\sigma$. When $\epsilon$ decreases this upward flow
moves towards the critical strip. For $\epsilon>2$  
the flow from the lower plane to the upper half-
plane takes mainly place outside the critical strip. Note that 
$\Psi_{e}(s,\epsilon,\varphi)$ has a pole at $s=1$ except for
$\epsilon=2,~\varphi=\pi$, which is a L-function. This shows that
a zero from the upward flow is located at
$s=1$ for these parameter values. For $\epsilon<2$  
the flow from lower to upper half-
plane takes mainly place inside the critical strip. 
When $\epsilon$
decreases the "train" of zeros moves thus to the left. Then 
zeros of $g_{5}(s)$ with $\sigma > 1/2$ 
are approached. If $\zeta(z_{\infty},1)\neq0$ these zeros
must be circumvented.  
The upward moving zeros can pass by these obstacles by a
bifurcation, i.e. merge
to a double zero, with the zero which form a circuit in the
clockwise sense around the zero, $z_{\infty}$, to $g_{5}(s)$. So 
for each simple zero $z_{\infty}$ with $\sigma > 1/2$ one extra
zero can be added to the upward flow.\\[1ex]
There
is however another possible obstacle for the moving zeros, namely
the famous hypothetical zeros, $z_{0}$, of $\zeta(s,1)$ outside
the critical line. 
The linearised motion around these zeros is
also of the form (18), except for a change of sign, so if $Re~z_{0}>1/2$
the circuits are now in the counter-clockwise sense. When the flow
of zeros comes in to the vicinity of such a zero, $z_{0}$, 
one zero leaves the "train" and starts to 
circulate around $z_{0}$. Thus, a
bifurcation is needed if $z_{0}$ is not a
common zero of $\zeta(s,1)$ and $g_{5}(s)$. 
The parameter values for the bifurcation that "creates" the zero
circulating around $z_{0}$ are below denoted by
$\epsilon_{B}$ and $\varphi_{B}$.
\\[2ex]
\section{Symmetric functions with $m > 5$}
Symmetric functions of the form (15) can 
be constructed for higher $m$ values. Here we concentrate on primes . For $m=7$ there are three
complex parameters and with the ansatz $\beta_{1}=1 + \epsilon
x_{1} e^{i\varphi},~\beta_{2}=1 + \epsilon x_{2} e^{i\varphi}$ and
$~\gamma=1 + \sqrt{7}e^{i2\varphi} + \epsilon x_{3} e^{i\varphi}$
the symmetry conditions lead to the following linear equation
for the real parameters $X=(x_{1}, x_{2},x_{3})$,
\beqar
\left(\begin{array}{ccc}
cos~\frac{4\pi}{7} &  cos~\frac{6\pi}{7} & \frac{1}{2} \\
cos~\frac{6\pi}{7} - \frac{\sqrt{7}}{2} &  
cos~\frac{2\pi}{7} & \frac{1}{2} \\
cos~\frac{2\pi}{7} &  cos~\frac{4\pi}{7} - \frac{\sqrt{7}}{2} 
& \frac{1}{2} \end{array} \right) X=0.
\eeqar{ax}
The rank, $r$, of this matrix is two and the null space is one
dimensional so there is only one possible construction, namely
\beq
\Psi_{e}(s,\epsilon,\varphi)=(1 +\frac{e^{i2\varphi}}{7^{s-1/2}})
\zeta (s,1) + \epsilon\frac{e^{i\varphi}}{7^{s}}g_{7}(s,x),
\eeq{msju}
where
\beq
g_{7}(s,x)~=~x_{1}(\zeta (s,2/7) +\zeta (s,5/7)) +
x_{2}(\zeta (s,3/7) +\zeta (s,4/7)) + x_{3}\zeta (s,1).
\eeq{h7s}   
Numerical investigations
of the corresponding matrices for all prime numbers $p<1000$ shows
that $p=4r \pm 1$ and the order of the matrices is $\frac{p-1}{2}$
so the dimension of the null space increases
with increasing $p$.\\[1ex]
An interesting case is $p=13$. Here the
null space is three dimensional so for each point $X$
on the sphere $S^{2}$ there is a symmetric function
\beq
\Psi_{e}(s,\epsilon,\varphi,X)=
(1 +\frac{e^{i2\varphi}}{13^{s-1/2}})
\zeta (s,1) + \epsilon\frac{e^{i\varphi}}{13^{s}}g_{13}(s,X).
\eeq{mtret}
Let us as in sect 3 assume the existence of simple Riemann
zeros, $z_{0}$, outside the critical line.
To each point $X$ there is one point $(\epsilon_{B},
\varphi_{B})$ where the bifurcation near $z_{0}$ ( see sect. 3)
takes place and for
$-X$ the corresponding point is $(\epsilon_{B},
\varphi_{B} + \pi)$. If now $X$ moves one circiut on the equator
of $S^{2}$ we expect that a closed curve $(\epsilon_{B}(X),   
\varphi_{B}(X))$ will be traced out. Let us assume that
$\epsilon_{B}=0$ does not lie on the curve. 
If then the equator is continously deformed to the north pole
, say, the closed curve must for some $X$ cross
$\epsilon_{B}=0$ which is impossible for simple zeros! The
situation is different for the bifurcations near $z_{\infty}$ since
the location of $z_{\infty}$ in the complex plane depends on $X$ and
$z_{\infty}$ is generally located on the critical line for some
$X$ values. Further work on these interesting symmetric functions is
in progress.

\begin{figure}
\begin{center}
\epsfig{file=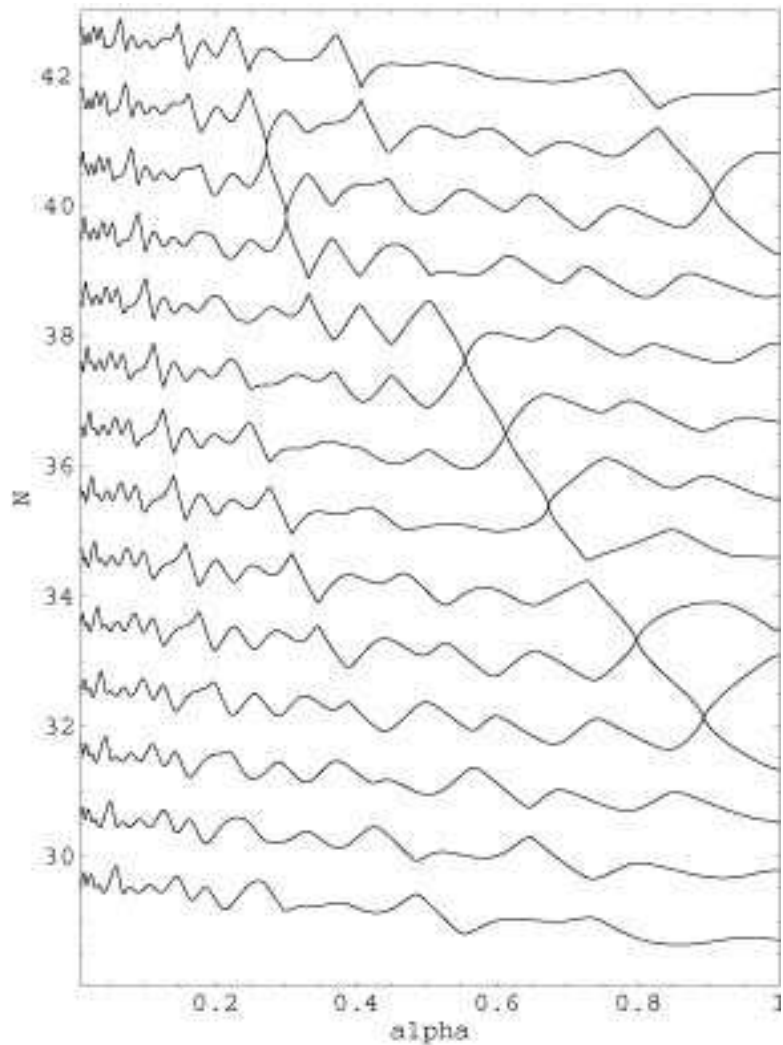}
\caption{The imaginary part of the 30:th to the
43:th zero of $\zeta(s,\alpha)$ in the range
$10^{-2}\leq\alpha\leq1$ is shown. Scaled values
,$N(t,\alpha)$
of $t=Im~z$ have been used. For this,
and all other figures, a 4:th order Runge-Kutta
method has been used to solve the equation of motion.
The suggested formula (5) for $N(T,\alpha)$ is
seen to work well. The
somewhat downsloping overall structure is probably due to
higher order terms in (5).} 
\end{center}
\end{figure}
\begin{figure}
\begin{center}
\epsfig{file=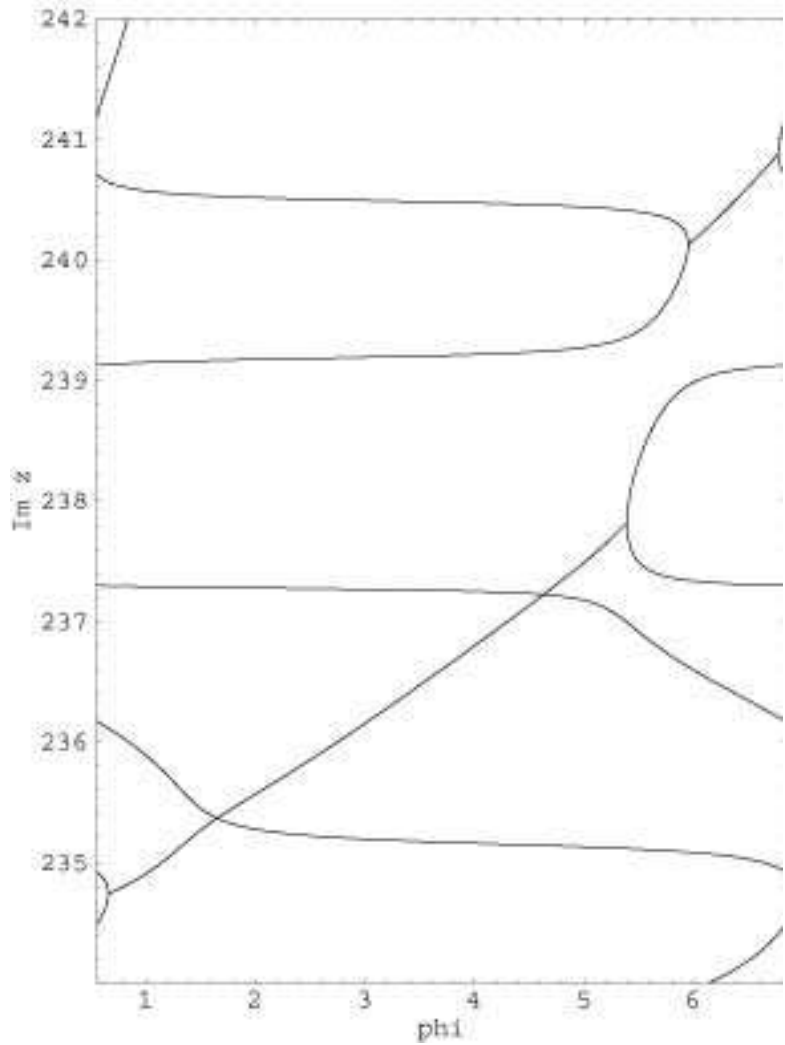}
\caption{A part of the spectra for the symmetric
function $\Psi_{5o}(s,\beta)$ in (7).
The parameter $\beta$ lies on the circle (8) and
an angle $\phi$ is used in the figure to parametrise
this circle. For $\beta=+i$ and $-i$, which corresponds
to left (and right) edge and $\phi\approx 5.73$ respectively,
$\Psi_{5o}$
are L-fuctions and all zeros for them are supposed to lie on the
critical line. Note the two upsloping segments which
join the double zeros on the critical line. To remedy
the singular equation of motion at the bifurcation points
a small complex constant has been added to the parameter $\phi$.}
\end{center}
\end{figure}

\begin{figure}
\begin{center}
\epsfig{file=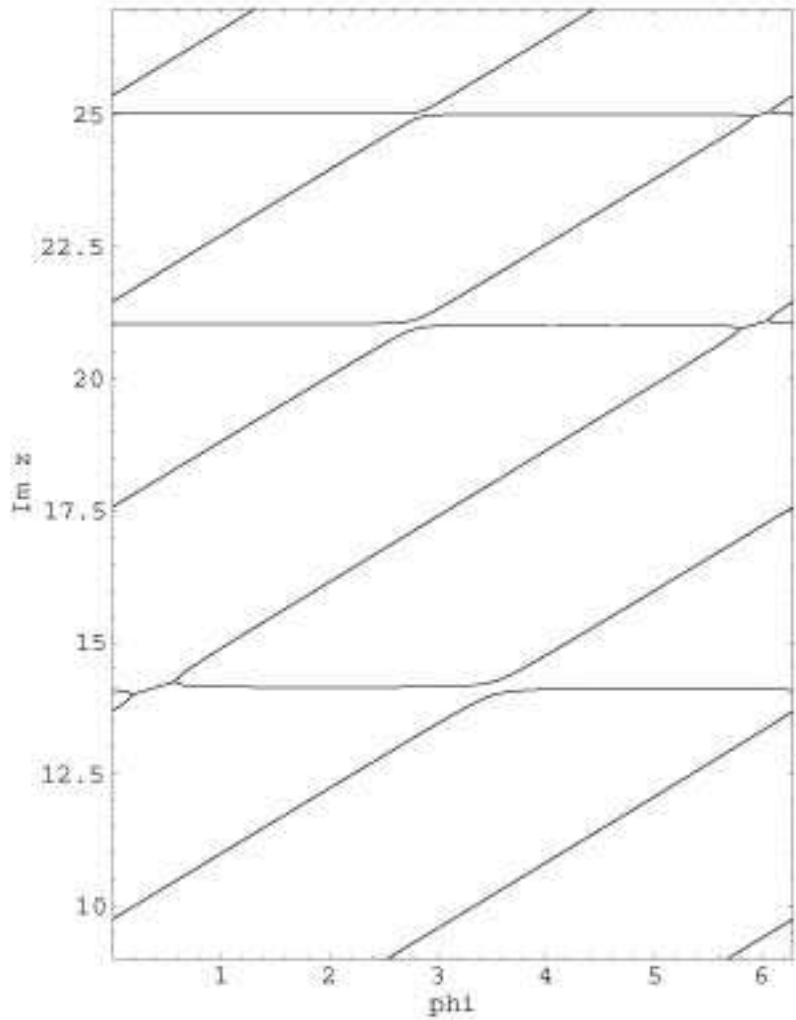}
\caption{A low lying part of the spectra for $\Psi_{e}(s,\epsilon,\varphi)$,
given by (15), is shown for $\epsilon=0.01$ and $0 \leq \varphi < 2 \pi$.
At $\epsilon=0$ the spectra
consists of upsloping trivial zeros crossing horisontal Riemann zeros.
Here these crossings are replaced by bifurcations
out in the complex plane or interactions on the critical line. As for
fig. 2 a small complex constant has been added to $\varphi$ to overcome
the bifurcations.}
\end{center}
\end{figure}

\end{document}